\documentclass[9pt,twocolumn,twoside]{osajnl}

\journal{ol} 

\setboolean{shortarticle}{true}

\usepackage{lineno}
\usepackage{amssymb}
\usepackage{braket}
\usepackage{epsfig,amsmath,graphics}
\usepackage{epstopdf}
\usepackage{float}
\usepackage{graphicx}
\usepackage{color}
\usepackage{xcolor}
\usepackage{gensymb}
\usepackage{natbib}
\usepackage{graphicx}
\usepackage[table]{xcolor}
\usepackage{slashed}
\usepackage{cancel}
\usepackage{multirow}
\usepackage{soul}
\begin{document}





\title{Watt-class injection-locked diode laser system at 399~nm for atomic physics} 
\author[1]{Rose Ranson}
\author[1]{Yifan Zhou}
\author[1]{Michael Hesford}
\author[1,2]{Jack Drouin}
\author[1]{Dhruv Azad}
\author[1]{Michalis Panagiotou}
\author[1,*]{Chris Overstreet}

\affil[1]{Department of Physics and Astronomy, Johns Hopkins University, Baltimore, Maryland, USA}

\affil[2]{JILA, NIST and University of Colorado, Department of Physics, Boulder, CO, USA}

\affil[*]{Corresponding author: c.overstreet@jhu.edu}

\ociscodes{(020.0020) Atomic and molecular physics; (120.0120) Instrumentation, measurement, and metrology; (140.0140) Lasers and laser optics.}

\date{Compiled \today}


\begin{abstract}
We demonstrate an injection-locked 399 nm laser system with up to 1 W output power and a locked power fraction of 0.57. The system consists of a high power, multimode diode laser that is seeded by $5$ mW from a single-mode external cavity diode laser.  The locked high-power laser inherits the frequency agility and linewidth of the seed laser with $3.9\ \text{kHz}$ broadening.  With active stabilization, the injection lock can be maintained for more than a day. We verify the utility of this system for atomic physics by performing spectroscopy of an ytterbium atomic beam. 
\end{abstract}
\maketitle

\section*{Introduction}

High-power, narrow-linewidth lasers are critical for atomic physics experiments. They enable laser cooling and magneto-optical trapping of atoms and are necessary for precision measurements \cite{Kim2025, hassan2025cryogenic, Morel2020, Parker2018, Asenbaum2020, Kim2020}, quantum simulations \cite{kaufman2016quantum, bernien2017probing, Takahashi2022}, and quantum information science \cite{deist2022mid, lis2023midcircuit}. It is challenging to achieve high-power lasers with narrow linewidths at blue or near-ultraviolet (UV) wavelengths. Existing technologies at these wavelengths include frequency-doubled lasers \cite{Tinsley2021}, solid-state lasers \cite{DeRose2023}, injection-locked single-mode lasers \cite{Saxberg2016, Hosoya2015, Zuo2025}, and injection-locked multimode lasers \cite{Pagett2016}. Near-UV lasers are becoming increasingly relevant for experiments that utilize alkaline-earth atoms \cite{Zhou2024, Abe2021} or other species with strong UV transitions \cite{vayninger2025magneto}. 

In this work, we use a $5.5\ \text{mW}$ seed laser 
to injection lock $57(1)$\%  of the power of a $1.2\ \text{W}$ ``follower'' multimode diode laser. 
The locked power is emitted at the frequency of the seed laser, here tuned to be the ${}^1S_0 \rightarrow {}^1P_1$ ($398.9\ \text{nm}$) transition of ytterbium, with a linewidth only $3.94(6)\ \text{kHz}$ broader than that of the seed. 
The remaining follower power is contained in a broad background spanning 1.3~THz.
We verify these results using heterodyne interferometry and direct atomic spectroscopy of a beam of cold ytterbium. 

\begin{figure}[t]
    \centering
    \includegraphics[width=1\linewidth]{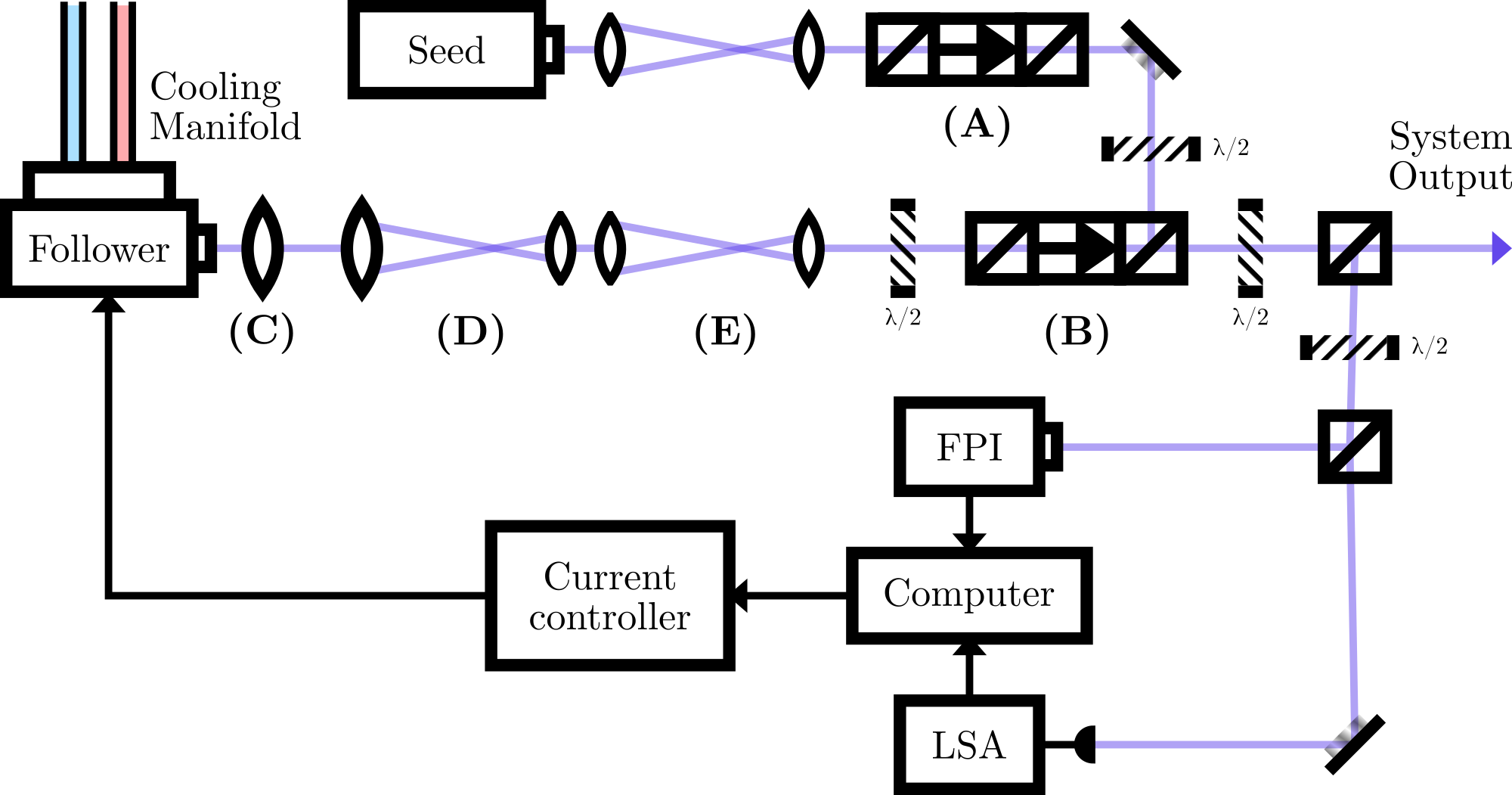}
    \caption{Schematic design of the injection-locked laser system. Faraday isolators \textbf{(A)} and \textbf{(B)}  (transmissivity 0.8, isolation efficiency $30\; \text{dB}$) prevent the follower laser from destabilizing the seed laser and prevent back-reflections from the system output from destabilizing the follower laser.  An initial aspheric lens \textbf{(C)} with focal length $13.9\ \text{mm}$ collects the highly divergent output of the follower diode. Telescopes \textbf{(D)} and \textbf{(E)} circularize and collimate the beam. The beam splitter before the system output sends $5\%$ of the power to a scanning Fabry-Perot interferometer (FPI) and laser spectrum analyzer (LSA), which monitor the follower beam and provide feedback to the current controller through a computer program. To stabilize the injection lock, a water-cooled manifold maintains the follower laser at $20\ \degree\text{C}$.}
    \label{fig:Optical_circuit}
\end{figure}

This system is at least 3 times more powerful than previously demonstrated injection-locked laser systems at 399~nm that are based on single-mode diodes \cite{GochnauerThesis, Hosoya2015} and can be utilized in applications, such as laser cooling and trapping, where single-mode operation is not essential.
Compared to alternatives that provide similar output power at near-UV wavelengths, the system is notable for its low cost and the ease of availability of its components. Since similar Watt-class multimode diodes are available at other wavelengths, analogous systems can likely be constructed at many wavelengths in the visible and near-UV spectrum.

\begin{figure}[t]
    \centering
    \includegraphics[width=1\linewidth]{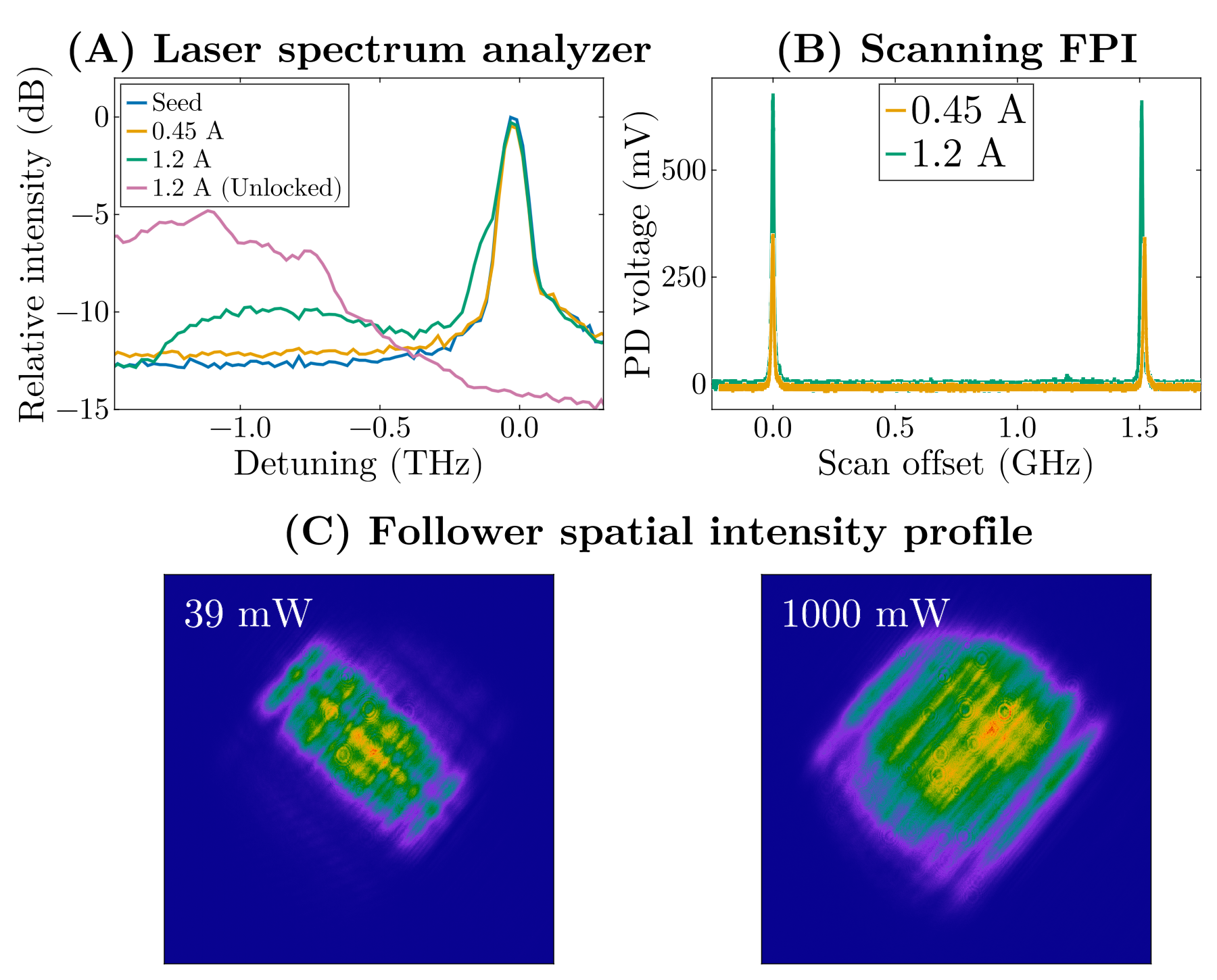}
    \caption{\textbf{(A)} Normalized power spectra of the seed laser (blue curve) and follower laser (orange, green, and purple curves) measured by the LSA at several follower currents. The unlocked follower spectrum (purple curve) is much broader ($\Gamma_\text{FWHM} =1.3\ \text{THz}$) than the resolution-limited locked follower spectrum ($\Gamma_\text{FWHM} = 70\ \text{GHz}$). \textbf{(B)} Transmission of the follower laser through the FPI with follower current 1.2 A (blue curve) and 0.45 A (orange curve). \textbf{(C)} Normalized spatial intensity profiles of the follower laser at the system output with 39 mW output power (left) and 1000 mW output power (right). Because the follower is a multimode laser, its profile significantly widens and changes shape as the current is increased.} 
    \label{fig:Currents_qualities_profiles}
\end{figure}

\section*{System configuration}

The setup for the injection lock is shown in Fig. \ref{fig:Optical_circuit}. The seed laser is an external-cavity diode laser (Toptica DL pro HP) that supplies $5.5\ \text{mW}$ of power at the follower diode facet. An initial Faraday isolator is used to prevent the seed laser from being destabilized by optical feedback from the follower laser. The seed light is overlapped with the follower light at the beam splitter of a second Faraday isolator. At the follower diode, the seed beam stimulates emission within its frequency bandwidth.

A multimode diode laser (Nichia NDV7975 in a Thorlabs LDM90 mount) serves as the follower. The beam emitted at the diode facet has a non-Gaussian spatial profile with divergence angles of $11$ degrees and $34$ degrees. This highly divergent beam is collected by an aspheric lens, and its astigmatism and ellipticity are then removed by a cylindrical telescope. The beam is then collimated by a circular telescope and passes through a Faraday isolator, which enables its overlap with the seed light and prevents back-reflections from the system output from reaching the follower diode. A small fraction ($5\%$) of the follower power is then split into a monitoring branch, where it is sent to a scanning Fabry-Perot interferometer (FPI) and a laser spectrum analyzer (LSA). The signals from these devices are provided to a computer program that adjusts the current controller (Thorlabs LDC4005) supplying the follower laser. We use a Thorlabs SA200-3B with a free spectral range of 1.5 GHz and finesse of 150 as the FPI and a Moglabs Compact Wavemeter as the LSA. The remaining follower beam power is sent to the system output. 

The quality of the lock is very sensitive to the angle and divergence of the seed light entering the follower diode. Initial alignment is performed by running the follower laser at its lasing threshold, where the lock quality is highest, and optimizing the positions and angles of the optics by using the LSA and FPI. This process is iterated as the follower power is raised to the desired level.

\subsection*{Stabilization and feedback}

An injection lock is achieved when the seed laser's frequency is within the lasing bandwidth of the follower diode and near a resonance of its cavity \cite{Lau2009}. The manufacturing process of the follower's Nichia NDV7975 diode leaves large variation in these diode-medium properties across different units. At ambient conditions of $25\ \degree\text{C}$, the diode used here possesses an unlocked spectrum centered at $402\ \text{nm}$. To maintain a lock at the primary $398.9\ \text{nm}$ (${}^1S_0 \rightarrow {}^1P_1$) transition and to improve stability, a water-cooled manifold was attached to the back of the diode mount to keep the diode at $20\ \degree\text{C}$. At this temperature, the central wavelength of the unlocked follower laser is shifted to 400 nm, reducing its frequency difference with the seed laser.

Once established, the lock typically ceases to function within a few minutes without active feedback because of ambient temperature drifts \cite{Liu2020, Saxberg2016}, so an active feedback mechanism \cite{Milanovic2025, Chen2021, Bordonalli1999} is necessary to achieve a long-term lock. Tuning the follower diode current changes the microscopic properties of the lasing medium, including its length, allowing the follower diode to be kept on resonance with the seed laser \cite{Saxberg2016}. The active feedback then consists of a computer program continuously making adjustments to the follower current to optimize the lock with respect to feedback metrics provided by the LSA and the FPI \cite{Saxberg2016}.

Coarse feedback is provided by the LSA (Fig. \ref{fig:Currents_qualities_profiles}(A)) which measures laser spectra across a $56\ \text{THz}$ range. As a metric of lock quality, we use the $L^2$ inner product of the normalized follower spectrum with the previously recorded seed spectrum. Since the linewidths of both the seed and locked follower are three orders of magnitude smaller than the resolution of the LSA, we find that the LSA alone does not provide a sufficiently precise feedback metric. 

Fine feedback is provided by the FPI (Fig. \ref{fig:Currents_qualities_profiles}(B)). Scanning the FPI's cavity length changes the frequency classes of the follower spectrum that can transmit through the cavity. When the unlocked follower laser is sent into the cavity, there is no change in transmitted power over the course of the scan because the spectrum has no features finer than the $1.5\ \text{GHz}$ free spectral range (see Fig. 2(A), purple curve). When the follower is locked, a significant fraction of its total power is localized near the frequency of the seed laser. Therefore, a transmission peak appears periodically as the cavity length is scanned. The maximum power transmitted over the course of the scan serves as the second feedback metric \cite{Saxberg2016}. 

To avoid locking to a suboptimal local maximum, the feedback program initially varies the follower current across a broad range and looks for the setpoint that maximizes the $L^2$ inner product with the seed spectrum, as measured by the LSA. The program then switches to optimizing the maximum power transmitted through the FPI using gradient descent. In this way, the injection lock has been realized and maintained under ambient temperature fluctuations of up to $2\ \degree\text{C}$ for over 24 hours.

\section*{Performance}

\begin{figure}[t]
    \centering
    \includegraphics[width=1\linewidth]{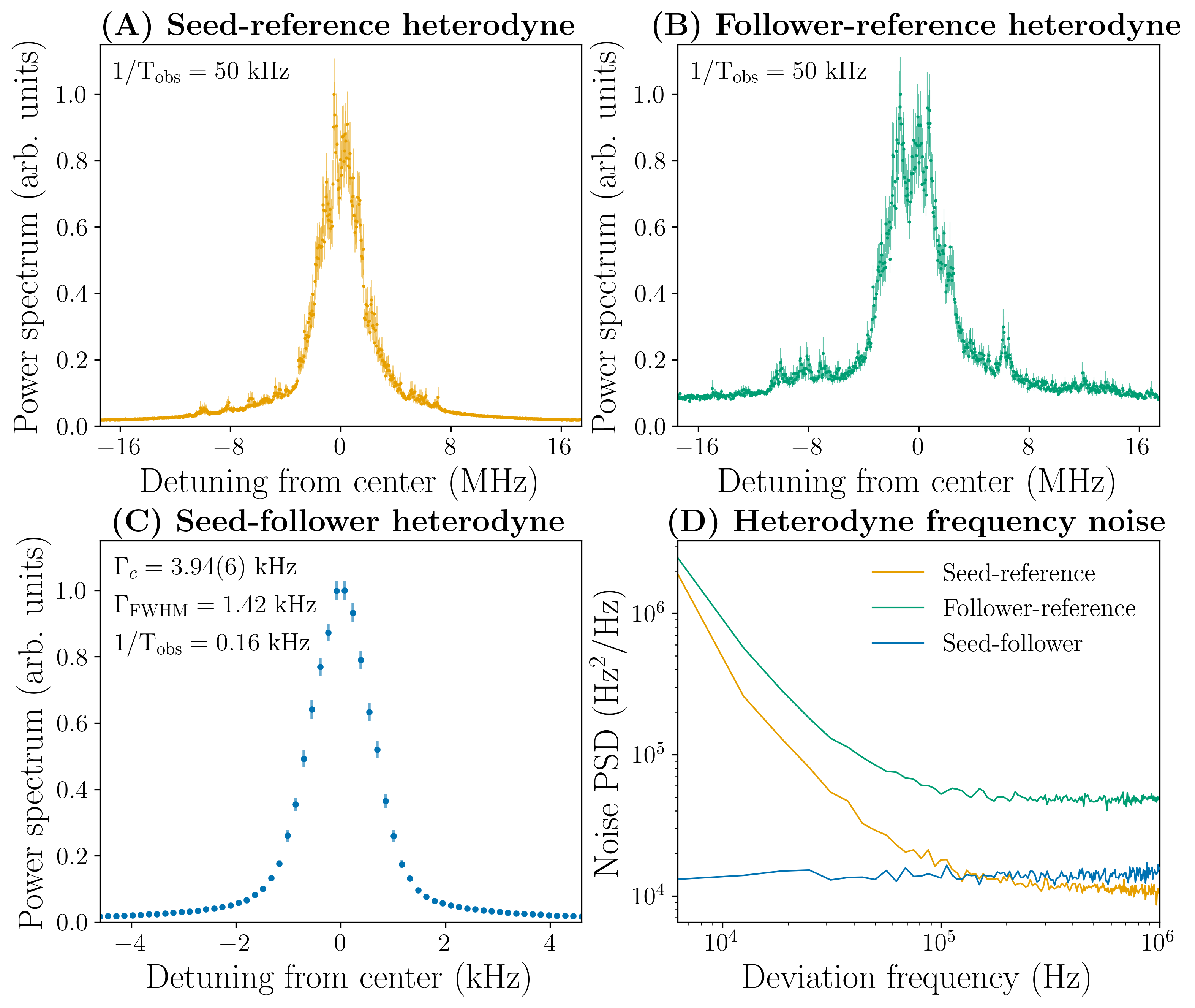}
    \caption{Heterodyne characterization of the seed and follower lasers. \textbf{(A, B)}  Heterodyne spectra of seed and follower laser, respectively, with a reference laser.  Each data set is collected over a time $T_\text{obs}$, and the plotted spectra represent the averages of several such data sets collected during a $1$~s time interval.  In (B), the  photodiode response to the bare follower beam is subtracted.  At this time scale, the seed and follower peaks have coherence linewidths of 8~MHz. \textbf{(C)}  Heterodyne spectrum of seed laser with follower laser, obtained by averaging data sets collected during a $10$~s time interval.  The peak has a coherence linewidth of $\Gamma_c = 3.94(6)\ \text{kHz}$ and a full width at half maximum $\Gamma_\text{FWHM} = 1.42\ \text{kHz}$. \textbf{(D)}  Frequency noise power spectral density of seed-reference heterodyne signal (orange curve), follower-reference heterodyne signal (green curve), and seed-follower heterodyne signal (blue curve).     
    }
    \label{fig:Heterodyning}
\end{figure}

\textbf{Linewidth}. 
Since the seed and follower spectra are narrower than the 10~MHz resolution of the FPI, a more precise technique is needed to determine their linewidths.
For this measurement, we perform heterodyne interferometry with a narrow-linewidth reference laser (Precilaser FL-SF-399-1.5-CW) and measure the interference signal with a $1\ \text{GHz}$ bandwidth photodiode. 
The reference laser is much narrower than the seed and follower lasers, so the heterodyne response is a direct characterization of their frequency spectra. The results of these measurements are shown in Fig. \ref{fig:Heterodyning}(A, B). The locked follower spectrum is qualitatively similar to the seed spectrum but possesses an additional broad noise background attributable to the unlocked power fraction. This broad background, which can be filtered if necessary \cite{Fasano2021}, is subtracted for the purpose of characterizing the spectrum of the locked power. 

To characterize differential broadening mechanisms between the follower and seed lasers, we perform heterodyne interferometry between them (Fig. \ref{fig:Heterodyning}(C)). The coherence linewidth between the follower and seed lasers, given by the inverse of the optical coherence time and calculated as suggested in \cite{VonBandel2016}, is $\Gamma_c = 3.94(6)\ \text{kHz}$, indicating that essentially all of the spectral width of the locked follower power is caused by phase noise inherited from the seed laser.

Finally, we compute the frequency noise power spectral density of each heterodyne measurement (Fig. \ref{fig:Heterodyning}(D)). The observed low-frequency drift of the seed laser is common to the follower laser, which does not add additional low-frequency noise between $10\ \text{kHz}$ and $100\ \text{kHz}$.

\textbf{Frequency agility}. The follower laser remains locked to the seed laser even as the seed frequency sweeps over a spectral range of $>2\ \text{GHz}$.

\textbf{Power output}. At a maximum current of $1.23\ \text{A}$, the follower laser emits $1.20\ \text{W}$
at the diode facet. At the system output, the maximum power is $958\ \text{mW}$.
The difference is primarily due to absorption losses within the Faraday isolator. However, only a fraction of this power is within the same frequency band as the seed spectrum, and the rest is part of a broad noise background. The locked power fraction represents the ratio of locked power to total power and characterizes the amount of power available for precision physics tasks. 

\begin{figure}[]
    \centering
    \includegraphics[width=1\linewidth]{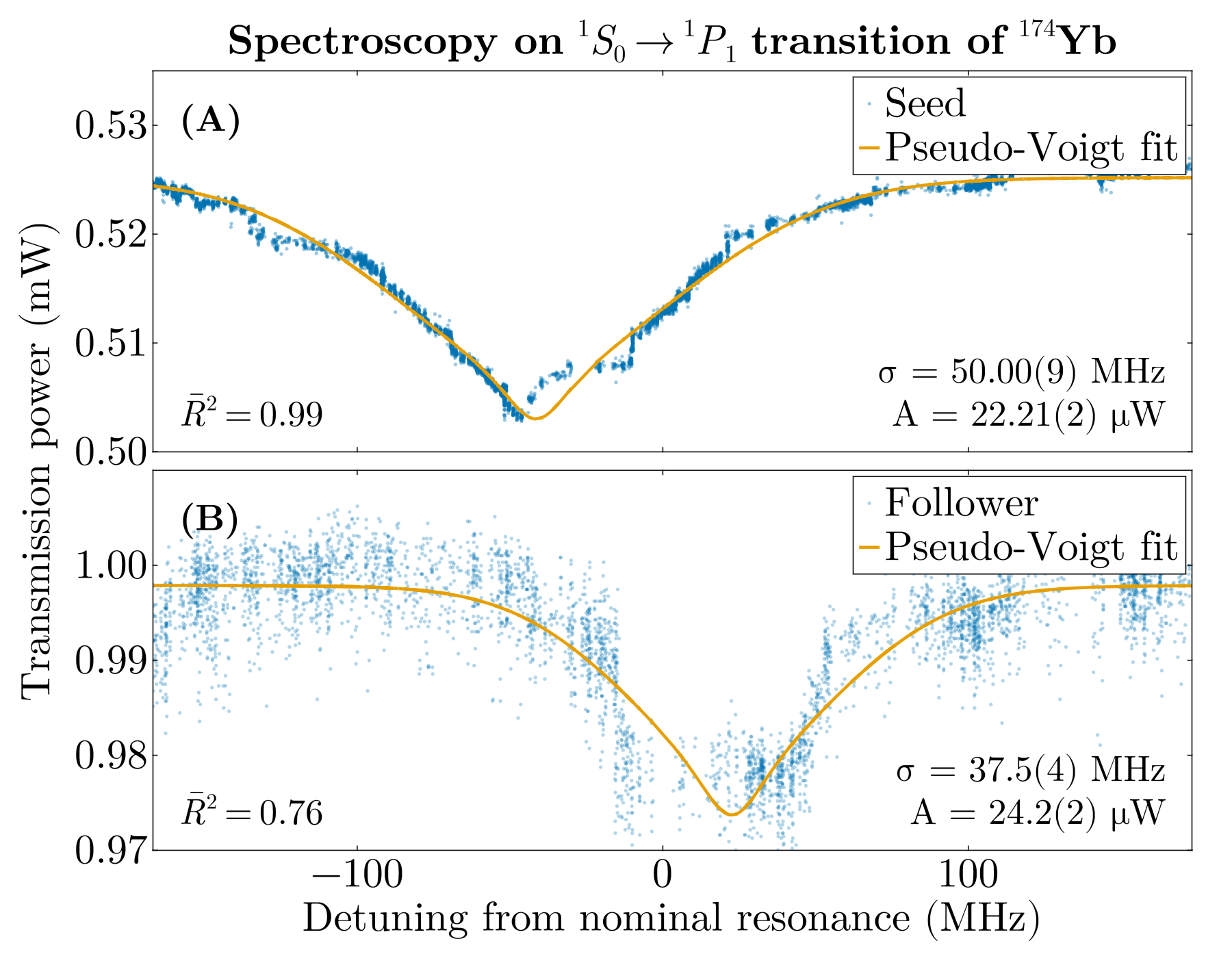}
    \caption{Transmission spectroscopy of a cold beam of ytterbium with \textbf{(A)} the seed and \textbf{(B)} the follower lasers. The yellow curves are pseudo-Voigt fits with Gaussian widths and amplitudes shown. The Lorentzian widths are fixed to the literature value of $29\ \text{MHz}$ \cite{Kroeze2025}. The resonances are shifted off their literature values in a manner consistent with a small Doppler shift arising from  non-orthogonality between the lasers and atomic beam. Adjusted $R^2$ values for each fit are also shown. Non-Lorentzian aspects of the line shapes are consistent with uncorrected power drifts in the seed and follower beams. }
    \label{fig:Transmission_spectroscopy}
\end{figure}

To measure the locked power fraction, we perform unsaturated transmission spectroscopy of an ytterbium atomic beam (AOSense Cold Atomic Beam Source) (Fig. \ref{fig:Transmission_spectroscopy}). As the linewidth of the primary ${}^1S_{0} \rightarrow {}^1P_1$ transition is $29\ \text{MHz}$ \cite{Kroeze2025} and much narrower than the unlocked power spectrum of the follower laser, the depth of the absorption feature for spectroscopy done with the follower laser is a proxy for the locked power fraction. 
From the ratio of optical depths measured with the seed and follower lasers, we calculate the locked power fraction to be 0.57(1), which implies a maximum locked power of $680(10) \ \text{mW}$ at the diode facet and 
$550(10)\ \text{mW}$
at the system output.
Coarser measures of the locked power fraction include the relative strengths of FPI transmission, the power difference measured by the LSA, and the relative amplitudes of the heterodyne resonances with the reference laser. We find each of these techniques to be in broad agreement.

The locked power fraction has a strong dependence on the follower current. At lower follower currents, the locked power fraction is larger because its foremost determinant is the ratio of seed power to follower power. When the follower diode current is increased, the total output power increases, but since the follower is a multimode diode laser, this is partially due to increased power emitted from the non-central spatial modes. An optimal injection lock alignment has maximal overlap with all of the modes, but individual non-central modes are optimally injection locked for slightly different alignments of the seed beam into the follower diode. For some alignments, we observe that the locked power fraction is non-monotonic as a function of follower current, which may indicate that the seed beam is non-centrally aligned.

The follower has $1\%$ relative intensity noise on a $1\ \text{s}$ time scale, which is $3.6$ times higher than the seed. This is sufficiently stable for laser cooling and trapping, but for applications where greater intensity control is required, the follower intensity could be actively stabilized.

\section*{Conclusion}

We have demonstrated a $398.9\ \text{nm}$ injection-locked multimode diode laser system with $550(10)\ \text{mW}$ maximum locked system output power and $3.94(6)\ \text{kHz}$ additional broadening compared to its seed laser. The remaining 
follower power at the system output is present as a broad ($1.3\ \text{THz}$) background. The system can be used to perform cold atomic beam spectroscopy and would be useful for laser cooling and magneto-optical trapping.

Such a system may find use in future experiments with neutral ytterbium, including quantum simulations \cite{Takahashi2022}, tests of the equivalence principle \cite{Hartwig2015}, and other searches for new physics \cite{Abe2021, Zhou2024}.  In addition, we anticipate that analogous systems could be realized at other wavelengths in the visible and near-UV spectrum.  Similar  Watt-class multimode diode lasers are available at many wavelengths between $375$~nm and $539$ nm.

\section*{Funding}

We acknowledge support from the National Science Foundation (FAIN 2409710).  M.P. acknowledges support from a William H. Miller III Graduate Fellowship. 

\section*{Disclosures}

The authors declare no conflicts of interest.

\section*{Data availability}

Data underlying the results presented in this paper may be obtained from the authors upon reasonable request.

\bibliographystyle{apsrev4-1}


%

\end{document}